\documentclass[twocolumn,showpacs,aps,prl,superscriptaddress]{revtex4}

\usepackage{graphicx}
\usepackage{amsmath}

\begin{document}

\title{The Two-flux Composite Fermion Series of Fractional Quantum Hall States in Strained Si}

\author{K. Lai}
\affiliation{Department of Electrical Engineering, Princeton
University, Princeton, New Jersey 08544}
\author{W. Pan}
\affiliation{Sandia National Laboratories, Albuquerque, NM
87185}
\author{D.C. Tsui} \affiliation{Department of Electrical
Engineering, Princeton University, Princeton, New Jersey 08544}
\author{S. Lyon}
\affiliation{Department of Electrical Engineering, Princeton
University, Princeton, New Jersey 08544}
\author{M. Muhlberger}
\affiliation{Institut fur Halbleiterphysik, Universitat Linz, Linz, Austria}
\author{F. Schaffler}
\affiliation{Institut fur Halbleiterphysik, Universitat Linz, Linz, Austria}

\date{\today}

\begin{abstract}

Magnetotransport properties are investigated in a high-mobility
two-dimensional electron system in the strained Si quantum well of
a (100) Si$_{0.75}$Ge$_{0.25}$/Si/Si$_{0.75}$Ge$_{0.25}$
heterostructure, at temperatures down to 30mK and in magnetic
fields up to 45T. We observe around $\nu=1/2$ the two-flux
composite fermion (CF) series of the fractional quantum Hall
effect (FQHE) at $\nu=2/3$, 3/5, 4/7, and at $\nu=4/9$, 2/5, 1/3.
Among these FQHE states, the $\nu=1/3$, 4/7 and 4/9 states are
seen for the first time in the Si/SiGe system. Interestingly, of
the CF series, the 3/5 state is weaker than the nearby 4/7 state
and the 3/7 state is conspicuously missing, resembling the
observation in the integer quantum Hall effect regime that the
$\nu=3$ is weaker than the nearby $\nu=4$ state. Our data indicate
that the two-fold degeneracy of the CFs is lifted and an estimated
valley splitting of $\sim$ 1K.

\end{abstract}
\maketitle

The composite fermion (CF) model\cite{jain,lopez,kalmeyer,halperin,perpectives,cf} 
has been very successful in explaining the principal fractional 
quantum Hall effect (FQHE) sequences of $\nu=p/(2p\pm1)$ 
($p$ = 1, 2, 3...) around the Landau level filling $\nu=1/2$. 
In this model, two fictitious flux quanta are attached to each electron. 
The so-formed new particles, composite fermions, can be treated as 
independent particles, and they move in a reduced effective magnetic 
($B$) field, $B_{eff}=B-2nh/e$. At $\nu=1/2$, $B_{eff}=0$ and the CFs 
form a Fermi sea. When $B_{eff}$ deviates from zero, Landau quantization
of the cyclotron orbits of CFs breaks up the CF energy continuum
into discrete Landau levels, giving arise to integer quantum Hall
effect (IQHE) states. The IQHE of the CFs at filling factor $p$
corresponds to the FQHE of the electrons at $\nu=p/(2p\pm1)$.

So far, research on CFs has mainly been carried out in the
two-dimensional electron system (2DES) in the lattice-matched
GaAs/Al$_x$Ga$_{1-x}$As heterostructures. Applicability of the CF
model in other material systems, especially the strained Si
quantum well in the Si/Si$_{1-x}$Ge$_x$ heterostructure, has not
been tested experimentally. In comparison to the GaAs/AlGaAs
systems, its multi-valley band structure, large electron effective
mass and $g$-factor, and negligible spin-orbit interaction are
expected to introduce new degrees of freedom into the CF physics.
Over the past ten years, the quality of strained Si has been
continuously improving\cite{xie,meyerson,sige} and a low-temperature 
2DES mobility as high as $\mu$ $\sim$ 600,000 cm$^2$/Vs has recently 
been reported\cite{okamoto}. However, high magnetic field measurements 
at dilution refrigerator temperatures are lacking and very few 
FQHE states have so far been observed. In fact, to date, only the 
two FQHE states at $\nu=2/3$ and 4/3 are firmly established
\cite{monroe,nelson,dunford,ismail,weitz}. Worse yet, the most prominent
$\nu=1/3$ FQHE state is still missing from observation and the
question whether the FQHE sequences in strained Si still follow
the two-flux CF series has thus remained unanswered.

In this paper, we report the magnetotransport data in a
high-mobility 2DES realized in the strained Si layer of a (100)
Si$_{0.75}$Ge$_{0.25}$/Si/Si$_{0.75}$Ge$_{0.25}$ quantum well
structure at temperatures ($T$) down to 30mK and in magnetic ($B$)
fields up to 45T. Around $\nu=1/2$, we observe the principal FQHE
states at $\nu=2/3$, 3/5, 4/7, and at $\nu=4/9$, 2/5, 1/3 --- the
two-flux CF series. This result demonstrates that the CF model
still applies in the Si/SiGe systems. Interestingly, of the
$p$/(2$p$$\pm$1) CF series, the 3/5 state is weak compared to the
nearby 4/7 state and the 3/7 state is conspicuously missing,
resembling the situation in the integer quantum Hall effect regime
where the $\nu=3$ state, due to the small valley splitting, is
weaker than the nearby $\nu=4$ state. In addition to the principal
FQHE states, weak $\rho_{xx}$ minima are also observed at
$\nu=4/5$ and $\nu=8/11$ around the even denominator $\nu=3/4$
fraction, as well as the $\nu=4/3$ and 8/5 FQHE states between
$\nu=1$ and 2. For $\nu$ $<$ 1/3, an insulating phase takes place
and no FQHE states are seen. The energy gap of the $\nu=1/3$ FQHE
state obtained from activation measurement is $\it
{\Delta}$$_{1/3}$ = 0.8K.

The specimen is an MBE grown, modulation doped n-type
Si$_{0.75}$Ge$_{0.25}$/Si/Si$_{0.75}$Ge$_{0.25}$ heterostructure.
The strained Si quantum well is 15 nm wide. Details of the growth
and the sample structure can be found in Ref. [9]. Ohmic contacts
to the 2DES were made by evaporating Au/Au:Sb and annealing at
370${^o}$C in a forming gas atmosphere. At $T\sim30$mK, the 2DES
has a density $n=2.7\times10^{11}$cm$^{-2}$ and mobility $\mu$ =
250,000cm$^2$/Vs, after low temperature illumination by a red
light-emitting diode. Standard low-frequency ($\sim7$Hz) lock-in
techniques were used to measure the magnetoresistivity $\rho_{xx}$
and the Hall resistivity $\rho_{xy}$.

Fig. 1 shows the $\rho_{xx}$ and $\rho_{xy}$ traces, taken at $T$
= 30 mK. In the low $B$ field regime, the Shubnikov-de Haas
oscillations are clearly resolved up to $\nu$ = 36. The odd
integer quantum Hall state, with an energy gap reflecting the
small valley splitting of the 2D electrons, appears at as early as
$\nu=11$. These features manifest the high quality of our
sample. Between $\nu=3$ and 4 and between $\nu=4$ and 5,
noticeable $\rho_{xx}$ dips are observed at $\nu=7/2$ and 9/2.
In the lowest Landau level between $\nu=1$ and 2, two FQHE
states, 8/5 and 4/3, are present. Similar to previous studies, the
$\nu=5/3$ state is missing\cite{dunford}.

\begin{figure}[!t]
\begin{center}
\includegraphics[width=3.2in,trim=0.3in 0.5in 0.2in 0.2in]{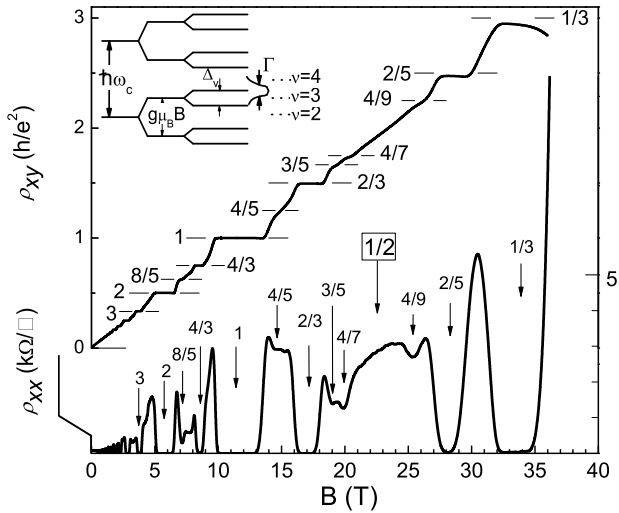}
\end{center}
\caption{\label{1}Diagonal resistivity $\rho_{xx}$ and Hall
resistivity $\rho_{xy}$ of the 2DES in a strained Si quantum well
at $T=30$mK. The 2DES has a density $n=2.7\times10^{11}$cm$^{-2}$
and mobility $\mu=250,000$cm$^2$/Vs. Major fractional quantum Hall
states are marked by arrows. The insert shows the electron Landau
level diagram. $\it \Gamma$ shows schematically the level
broadening.}
\end{figure}

For $\nu<1$, the FQHE states are observed at $\nu$ = 1/3, 2/5,
4/9, 4/7, 3/5, and 2/3. Weak $\rho_{xx}$ minima are also observed
at $\nu=4/5$ and 8/11. Evidence of the $\nu$ = 2/5 FQHE state was
first reported by Dunford $\it et$ $\it al.$ in Ref. [13].
However, only in this high quality sample the formation of the
$\nu=2/5$ FQHE state is unambiguously established: Its $\rho_{xx}$
is virtually zero within our experimental resolution and
$\rho_{xy}$ shows a well-developed plateau. The $\nu=4/7$ and
$\nu=4/9$ states are seen for the first time. Interestingly, while
the $\nu=4/7$ and 4/9 states show well-developed $\rho_{xx}$
minima, the $\nu=3/5$ state is weaker than the nearby $\nu=4/7$
FQHE states and the $\nu=3/7$ state is missing. When $\nu<1/3$,
the sample has entered into an insulating phase and no FQHE states
are observed.

The most striking feature in this figure is the observation of a
well-developed FQHE at $\nu=1/3$. To our knowledge, this is the
first time that the $\nu=1/3$ state has been observed in the
Si/SiGe system. In $\rho_{xx}$, the minimum has reached a low
value of $\sim50\Omega$/square, and the temperature dependent
measurements show that the $\rho_{xx}$ is thermally activated. In
$\rho_{xy}$, an apparent plateau accompanies the $\rho_{xx}$
minimum. We note that its quantization, however, is not exact.
Since our sample is not patterned into a Hall bar, contact
misalignment is likely to happen and the final value of
$\rho_{xy}$ may be contaminated with a component of $\rho_{xx}$.
The usual method to show and eliminate this $\rho_{xx}$ mixing
effect is to reverse the $B$ field direction, which,
unfortunately, could not be done with the hybrid magnet in our
experiment. The mixing from the strongly increasing $\rho_{xx}$,
as the insulating phase is approached, gives rise to the
continuous drop of $\rho_{xy}$ when the $B$ field is further
increased beyond $\nu=1/3$.

Fig. 2a shows the temperature dependence of $\rho_{xx}$ for
$\nu<1$. In Fig. 2b the $\rho_{xx}$ minima at $\nu=1/3$, 2/5, 4/9,
and 4/7 are plotted as a function of 1/$T$ on a semi-log scale.
For all four states, $\rho_{xx}$ shows the thermally activated
behavior, as expected, and the energy gaps for the $\nu=1/3$ and
2/5 states, obtained by fitting the linear portion of the data to
$\rho_{xx}\propto$ exp(-$\it{\Delta}$/2k$_B$T), are $\it
{\Delta}$$_{1/3}\sim$ 0.8K and $\it {\Delta}$$_{2/5} \sim$ 1.3 K.
(For the $\nu=4/9$ and $\nu=4/7$ states, the linear range of the
data is small and the estimated gaps are $\it
{\Delta}$$_{4/9}\sim$ $\it {\Delta}$$_{4/7}\sim$ 0.2K.) Unlike in
GaAs, here, $\it {\Delta}$$_{1/3}$ is smaller than $\it
{\Delta}$$_{2/5}$. We believe that this difference is due to the
close proximity of the 1/3 state here to the high $B$ field
insulating phase.

\begin{figure}[!t]
\begin{center}
\includegraphics[width=3.2in,trim=0.2in 0.5in 0.4in 0.2in]{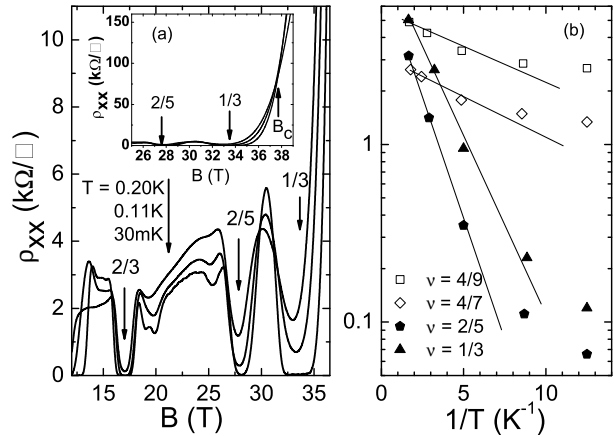}
\end{center}
\caption{\label{2}(a) $\rho_{xx}$ vs. $B$ at selected temperatures. 
The inset shows the transition from the $\nu=1/3$ FQHE state to 
the high $B$ insulating state. The critical $B$ field is indicated 
by the arrow marked Bc and the critical resistivity is 
$\rho_{xxc}\sim80$k$\Omega$/square. (b) Temperature dependence 
of $\rho_{xx}$ minima of the FQHE states at $\nu=1/3$, 2/5, 4/9, 
and 4/7. Lines are linear fits to the data points and the 
energy gaps $\it\Delta$$_{4/9}\sim\it\Delta$$_{4/7}\sim0.2$K, 
$\it\Delta$$_{1/3}\sim$ 0.8K and $\it\Delta$$_{2/5} \sim$ 1.3K.}
\end{figure}

The two sequences of FQHE states evolving toward $\nu=1/2$ from
$\nu=2/3$, 3/5, and 4/7, and from $\nu=1/3$, 2/5, and 4/9 follow
the two-flux composite fermion series of $\nu=p/(2p\pm1)$. It
shows that the CF model still applies in our multi-valley strained
Si system. In Fig. 3, we compare the $\rho_{xx}$ data around
$\nu=1/2$ (or $B_{eff}=0$) and that around $B=0$. It is clear that
there exits a one-to-one correspondence between the IQHE states
and FQHE states --- the positions of the  $\rho_{xx}$ minima and
their relative strengths --- for the same value of $\nu$ and $p$.
For example, the wide range of vanishing $\rho_{xx}$ in the
$\nu=1/3$ FQHE state corresponds to a similar range of vanishing
$\rho_{xx}$ in the $\nu=1$ IQHE state. A relatively weaker
$\nu=2/5$ FQHE state corresponds to a less extended vanishing
$\rho_{xx}$ region of the $\nu=2$ IQHE state. More interestingly,
the weak $\nu=3/5$ state (the missing $\nu=3/7$ state) when
compared with the nearby $\nu=2/3$ and 4/7 states ($\nu=2/5$ and
4/9 states) resembles what is seen between $\nu=2$ and $\nu=4$
where the $\nu=3$ minimum is also the weakest. This similarity
also appears in the temperature dependence (not shown). The fact
that the $\nu=3/5$ FQHE state is more easily destroyed by raising
the sample temperature than the $\nu=4/7$ state resembles again
what has been observed in the IQHE regime, where the $\nu=3$ state
also is more easily destroyed than the $\nu=4$ state. For the two
weak minima at $\nu=4/5$ and 8/11, they correspond to the FQHE
states of electrons at $\nu=4/3$ and 8/5. On the other hand, these
two states can also be mapped to the IQHE states at $p=1$ and 3 of
the CFs with four flux quanta forming at $\nu=3/4$\cite{yeh}.

\begin{figure}[!t]
\begin{center}
\includegraphics[width=3.2in,trim=0.2in 0.3in 0.4in 0.2in]{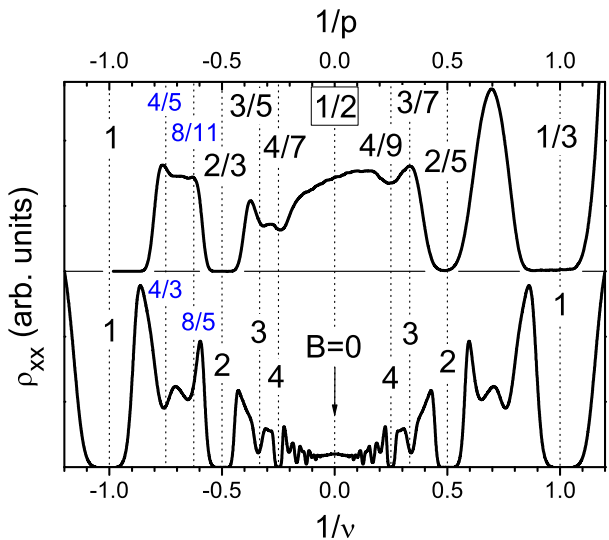}
\end{center}
\caption{\label{3}Comparison of the two-flux CF series around 
$\nu=1/2$ and the IQHE states around $B=0$. The IQHE trace was 
taken at an elevated temperature of $T\sim0.3$K. The x-axis 
is 1/$\nu$ for electrons and 1/$p$ for CFs. }
\end{figure}

The similarity between the IQHE states and the FQHE states has
long been seen in the GaAs/AlGaAs system and was, in fact, the
primary motivation for the composite fermion model. It maps the
FQHE of electrons onto the IQHE of CFs and thus naturally explains
the one-to-one correspondence in the B field positions of
$\rho_{xx}$ minima. Moreover, the CF model is successful in
explaining the relative strength of the FQHE states by relating
their many-body energy gaps to the single-particle CF Landau level
separations. Consequently, the larger $p$ state, corresponding to
a smaller CF Landau level separation, is generally weaker.
However, it is not obvious whether this simple explanation still
applies in the presence of other degree of freedom, for example,
in our case the valley degeneracy. In the following, we show that,
although a little bit more complicated, the relative strength of
the FQHE in the strained Si can indeed be understood from the
single-particle CF Landau level diagram.

First, we recall the physical origin of why the $\nu=3$ IQHE state
is weaker than the nearby $\nu=2$ and 4 states. In strained (100)
Si, where the two conduction valleys perpendicular to the surface
are lower in energy, the strength of an IQHE state is determined
by four energy scales --- the Landau level separation
($\hbar\omega_c$), its level broadening($\it \Gamma$), the Zeeman
splitting ($g\mu_BB$), and the valley splitting ($\it
{\Delta}$$_v$). The magnitude of $\it {\Delta}$$_v$ is
particularly important to the odd IQHE state. As shown
schematically in the inset of Fig. 1, the energy gap of an odd
IQHE state depends directly on $\it {\Delta}$$_v$. Since the
valley splitting is generally smaller than the Landau level
separation and Zeeman splitting, this explains why the $\nu=3$
state is weaker than the nearby $\nu=2$ and $\nu=4$ state. To
estimate $\it {\Delta}$$_v$, we use the fact that the odd IQHE
state appears when $\it{\Delta}$$_v \geq\it {\Gamma}$. For our
specimen, $\it{\Gamma}=\hbar/\tau=\hbar e/m \tau\sim$ 0.3K, where
$\tau$ is the transport lifetime, $m=0.2m_e$ is the effective mass
of the electron, and $\mu=250,000$cm$^2$/Vs. Based on our
observation that the first resolved odd integer state is $\nu=11$,
we estimate $\it{\Delta}$$_v\sim0.3$K at $B=1.0$T.

In the FQHE regime, CFs form at $\nu=1/2$\cite{jain,lopez,kalmeyer,
halperin,perpectives,cf}. When B deviates from $B_{1/2}$ (the $B$ 
field at $\nu=1/2$), the CFs see a reduced effective $B$ field, 
$B_{eff}=B-B_{1/2}$, and form Landau levels with a level separation 
of $\hbar B_{eff}/m^*$ ($m^*$ is the CF effective mass), giving rise 
to the IQHE of CFs. The IQHE state at Landau filling $p$ of the CFs 
corresponds to the FQHE state of electrons at the filling $\nu=p/(2p\pm1)$. 
To draw the CF Landau level diagram, again, we need to know the 
four energy scales --- the CF Landau level separation ($\hbar\omega_c^*$), 
its broadening ($\it\Gamma^*$), the Zeeman splitting ($E_z^*$), 
and the valley splitting ($\it\Delta$$_v^*$). To estimate the Landau level
separation, we use the effective mass, $m^*\sim1.4m_e$, obtained
from a scaling argument\cite{yeh,mass,du1}. At $p=3$ (or $\nu=3/5$ and 3/7),
$\hbar\omega_c^*\sim3$K. The CF Landau level broadening
$\it{\Gamma}^*=\hbar e/m^*\mu^*\sim1$K, where $\mu^*$ is the CF
mobility. We calculate it from the resistivity at $\nu=1/2$, using
$\mu^*=1/ne\rho_{xx}(\nu=1/2)\sim7.5\times10^3$cm$^2$/Vs. As to
the CF Zeeman splitting, we recall that the effective $g$-factor
of the CF is the same as that of the electron\cite{du2}. Consequently,
$E_z^*=g^*\mu_BB\sim30$K. Since $E_z^*\gg\hbar\omega_c^*$, the
spin-degeneracy of the CF Landau level is completely lifted and
the CFs can be viewed as spinless. Finally, from the similarity in
$\rho_{xx}$ between the CFs and the electrons, we conclude that at
$p=3$ the two-fold valley degeneracy is also lifted for the CFs.
Following the same criterion that  $\it \Delta$$_v^*\sim\it \Gamma^*$
for the first resolved odd IQHE state of the CFs, we estimate a
valley splitting $\it\Delta$$_v^*\sim1$K.

Finally, a close examination of the temperature dependence of
$\rho_{xx}$ reveals a transition from the FQHE state at $\nu=1/3$
to an insulating state at higher $B$ fields, as shown in the inset
of Fig. 2a. The critical $B$ field, at which the $\rho_{xx}$ is
temperature independent, is $B_c\sim37.6$T, and the critical
resistivity is $\rho_{xxc}\sim80$k$\Omega$/square. This
$\rho_{xxc}$ is considerably larger than
$h/e^2\sim26$k$\Omega$/square, observed in GaAs for the same
transition\cite{shahar}.

To summarize, we have observed around $\nu=1/2$ the principal FQHE
states, corresponding to the two-flux CF series $p/(2p\pm1)$, at
$\nu=2/3$, 3/5, 4/7, and at $\nu=4/9$, 2/5, 1/3 in a high-mobility
two-dimensional electron system in the strained Si layer of a
(100) Si$_{0.75}$Ge$_{0.25}$/Si/Si$_{0.75}$Ge$_{0.25}$ quantum
well structure. The strengths of the two-flux CF series display a
striking resemblance to those of the IQHE states of electrons,
indicating that the two-fold degeneracy of the CF is lifted and an
estimated valley splitting of $\sim1$K. Our results show that the
CF model still applies to the multi-valley Si/SiGe system.

This research was supported by the DOE and the NSF. The work at
the NHMFL is supported by NSF Cooperative Agreement No.
DMR-9527035 and by the State of Florida. Sandia is a multiprogram
laboratory operated by Sandia Corporation, a Lockheed-Martin
company, for the U.S. Department of Energy under Contract No.
DE-AC04-94AL85000. We thank E. Palm, T. Murphy, S. Hannahs, B.
Brandt, and the Hybrid Magnet operation team for their
experimental assistances.


\begin{thebibliography}{90}
\bibitem{mass} Since the CF effective mass $m^\ast$ is determined by 
electron-electron interaction, it scales as $B_{1/2}$. In the 
GaAs/AlGaAs system, it was found that $m^\ast/m_e\sim0.26\times(B_{1/2})^{1/2}
=3.4\times(B_{1/2})^{1/2}/\epsilon_{GaAs}$\cite{yeh, du1}, 
where $B_{1/2}$ is in units of Tesla and $\epsilon_{GaAs}$ 
is the dielectric constant of GaAs. Replacing $\epsilon_{GaAs}$ 
with $\epsilon_{Si}=11.7$ and $B_{1/2}\sim22.5$T, we find 
$m^\ast\sim1.4m_e$.
\end{thebibliography}
\end{document}